\documentclass[%
 prd,
 reprint,
superscriptaddress,
preprintnumbers,
nofootinbib,
 amsmath,amssymb,
 aps,
]{revtex4-1}
\pdfoutput=1
\usepackage{hyperref}

\usepackage[textwidth=1.25cm,textsize=footnotesize,
]{todonotes}

\usepackage[utf8]{inputenc}

\newcommand{\be}{\begin{equation}}
\newcommand{\ee}{\end{equation}}
\newcommand{\ba}{\begin{eqnarray}}
\newcommand{\ea}{\end{eqnarray}}

\renewcommand{\l}{\left(}
\renewcommand{\r}{\right)}

\begin{document}

\preprint{INR-TH/2016-004}

\title{BEST sensitivity to {\cal O}(1)\,eV sterile neutrino}
%
\author{Vladislav Barinov}
\email{barinov.vvl@gmail.com} 
\affiliation{Institute for Nuclear Research of the Russian Academy of Sciences,
  Moscow 117312, Russia}
\affiliation{Physics Department, Moscow State University, 
Vorobievy Gory, Moscow 119991, Russia}

\author{Vladimir Gavrin}
\email{gavrin@inr.ru}
\affiliation{Institute for Nuclear Research of the Russian Academy of Sciences,
  Moscow 117312, Russia}

\author{Dmitry Gorbunov}
\email{gorby@ms2.inr.ac.ru}
\affiliation{Institute for Nuclear Research of the Russian Academy of Sciences,
  Moscow 117312, Russia}
\affiliation{Moscow Institute of Physics and Technology, 
  Dolgoprudny 141700, Russia}

\author{Tatiana Ibragimova}
\email{tvi@inr.ru}
\affiliation{Institute for Nuclear Research of the Russian Academy of Sciences,
  Moscow 117312, Russia}


\begin{abstract}
Numerous anomalous results in neutrino oscillation experiments can be
attributed to interference of $\sim 1$\,eV sterile neutrino. 
The specially designed to fully explore the Gallium 
anomaly Baksan Experiment on Sterile Transitions 
(BEST) starts next year. 
We investigate the sensitivity of BEST in searches for sterile
neutrino mixed with electron neutrino. Then, performing the combined
analysis of all the Gallium experiments (SAGE, GALLEX, BEST) 
we find the regions in model
parameter space (sterile neutrino mass and mixing angle), which 
will be excluded if BEST agrees with no sterile neutrino hypothesis. 
For the opposite case, if BEST observes the signal as it follows from
the sterile neutrino explanation of the Gallium (SAGE and GALLEX)
anomaly, we show how BEST will improve upon the present estimates of the 
model parameters. 
\end{abstract}

\maketitle

\section{Introduction}
\label{sec:Intro}

Neutrino oscillations\footnote{There are antineutrino oscillations as
  well, but in the general discussion we do not distinguish the two
  cases.} provide with the only direct irrefutable evidence for
incompleteness of the Standard Model of particle physics
(SM). Moreover, while most issues of the neutrino experiments can be
properly addressed by making at least two out of three SM neutrinos
massive, there are several {\it anomalous} results which are
definitely beyond the grasp of this simple extension. 

The results
of experiments LSND\,\cite{Athanassopoulos:1995iw,Aguilar:2001ty}, 
MiniBooNE\,\cite{AguilarArevalo:2010wv,Aguilar-Arevalo:2013pmq}, 
SAGE\,\cite{Abdurashitov:1998ne,Abdurashitov:2005tb},  
GALLEX\,\cite{Kaether:2010ag}, 
analysis of measured reactor antineutrinos\,\cite{Mueller:2011nm,Huber:2011wv} 
show \emph{anomalous change of neutrino fluxes}. If attributed to
oscillations, it requires much bigger values of neutrino squared mass
difference, $\Delta m^2_{anom}\simeq 1 $\,eV$^2$, 
as compared to the already known values of the two mass squared differences 
(so called solar, $\Delta m^2_{sol}\approx 7.5\times 10^{-5}$\,eV$^2$ 
and atmospheric, 
$\Delta m^2_{atm}\approx 2.5\times 10^{-3}$\,eV$^2$
\cite{Agashe:2014kda})   
sufficient to explain the results of the great majority of neutrino
oscillation experiments. The hierarchy between the two mass
differences, $\Delta m^2_{sol}\ll\Delta m^2_{atm}$,  
can be described by three neutrino eigenstates and hence is consistent 
with three neutrino pattern. The third mass difference making the pronounced
hierarchy  $\Delta m^2_{sol} \ll \Delta m^2_{atm}\ll \Delta
m^2_{anom}$ asks for (at least) one more neutrino eigenstate,
that the SM does not have. A hunt for the new light neutrino specie is
the main task of many developing projects\,\cite{Abazajian:2012ys}.

The anomalies naturally form two classes. There are anomalous
appearances (excesses of signal events) and anomalous disappearances
(lacks of signal events). 
An anomalous disappearance of neutrinos can point at oscillations into the
(hypothetical) fourth light neutrino state, singlet with respect to
the SM gauge group, and hence called {\it sterile}, while the SM
neutrinos are dubbed {\it active}. In particular, observed by
experiments SAGE\,\cite{Abdurashitov:2009tn}  
and GALLEX\,\cite{Kaether:2010ag} lack of electron neutrinos from artificial
radioactive neutrino sources can be explained by oscillations into sterile
neutrinos, which obviously escape detection. Then electron neutrino
flux measured at a distance $r$ from the source is proportional to the
electron neutrino {\it survival probability} (against transition 
into the sterile state). For the artificial
sources under discussion, the neutrino flux is
quasi-monochromatic. The survival probability for neutrino of energy
$E_\nu$ is determined in the two-neutrino effective oscillating system
through the sterile-active mixing angle $\theta$ and the squared mass
difference $\Delta m^2$ (saturated mostly by the sterile neutrino 
squared mass) as follows, see e.g.\,\cite{Bilenky:1998dt},  
\begin{equation}
P\l E_\nu,r\r=1-\sin^22\theta\,\sin^2\!\!\l \!\! 1.27\times 
\frac{\Delta m^2[\text{eV}^2] 
\,r[\text{m}]}{E_\nu[\text{MeV}]}\!\r\!.
\label{survival}
\end{equation}
The joint analysis\,\cite{Giunti:2010zu} 
of the four Gallium anomalous results (two per each
experiment) reveals the anomaly---disappearance of electron
neutrinos---at the statistical level of 2-3 standard deviations. 
Within the hypothesis of
oscillations into sterile neutrino, the best fit of model parameters
entering the survival probability\,\eqref{survival} have typical 
values in the region\,\cite{Giunti:2012tn} 
\begin{equation}
\label{best-fit-region}
\Delta m^2\sim 2\,\text{eV}^2\,,\;\;\;\;\sin^22\theta\sim0.3\!-\!0.5\,.
\end{equation}

The new Gallium experiment 
BEST\,\cite{Gavrin:2010qj,Gavrin:2011zz,Abazajian:2012ys}  
in Baksan Neutrino Observatory was
proposed to thoroughly explore the Gallium anomaly. It is
a short-baseline experiment utilizing the 
artificial compact ${}^{51}$Cr source of almost monochromatic 
electron neutrinos to be measured at effectively two distances of
$\sim0.4$\,m and $\sim0.8$\,m from the source. 
After accurate measurement and detailed analysis of the 
neutrino-Gallium cross
section\,\cite{Frekers:2011zz,Frekers:2013hsa,Frekers:2015wga} it has
been recently approved, see Refs.\,\cite{Abdurashitov:2011zz,Gorbachev:2012ka,Gorbachev:2013bya,Gavrin:2013uoa,
Gavrin:2015aca} for description of the passed and present R\&D
stages. It will start supposedly next year with production of artificial
3\,MCi radioactive ${}^{51}$Cr source of electron neutrinos. In this
paper we refine the preliminary estimates \cite{Abazajian:2012ys} of
the BEST sensitivity to the sterile neutrino parameters and its prospects
in exploring the Gallium anomaly. 

The paper is organized as follows. Sec.\,\ref{sec:Exp} contains
brief descriptions of the main idea of the experiment, the
artificial source, detecting technique, data processing, 
registration efficiency, and final accuracy in measurement of the
electron neutrino flux. 
We discuss the anomalous results of the Gallium experiments
in sec.\,\ref{sec:Anomaly} and  
 present the region in the
sterile neutrino model parameter space ($\Delta m^2$,\,$\sin^22\theta$)
favored by the Gallium anomaly. 
In sec.\,\ref{sec:Results} we 
outline the regions to be excluded by BEST, if its result is
consistent with no oscillation hypothesis, and the regions to be
excluded by the joint analysis of all the Gallium
experiments. Likewise we consider the possibility that the BEST future
result is consistent with predictions of the sterile neutrino model
with parameters tuned at the best-fit to the Gallium anomaly, 
outline the favored by BEST region in this case and
present the region chosen by the joint analysis of all the
Gallium experiments.  We summarize in sec.\,\ref{sec:Concl}.

\section{Description of the experiment: layout, operation and data analysis}
\label{sec:Exp}

BEST is a short-baseline neutrino oscillation experiment. The size is
determined from eq.\,\eqref{survival} by the best-fit values
\eqref{best-fit-region}, which for MeV-scale neutrino energy implies
1\,m-scale oscillation length. The artificial radioactive source of
3MCi is made of ${}^{51}$Cr, which decays emitting quasi-monochromatic
neutrinos of energies $E_{1a}=0.747$\,MeV (dominant mode), 
$E_{1b}=0.752$\,MeV, $E_{2a}=0.427$\,MeV, and $E_{2b}=0.432$\,MeV. 
The source intensity is measured with not worse  
than 0.5\% accuracy by making use of the
methods presented in 
Refs.\,\cite{Gorbachev:2014my1,Veretenkin:2014my1,Gorbachev:2015my1}.  

\begin{figure}[!htb]
\centering
\includegraphics[width=0.9\columnwidth]{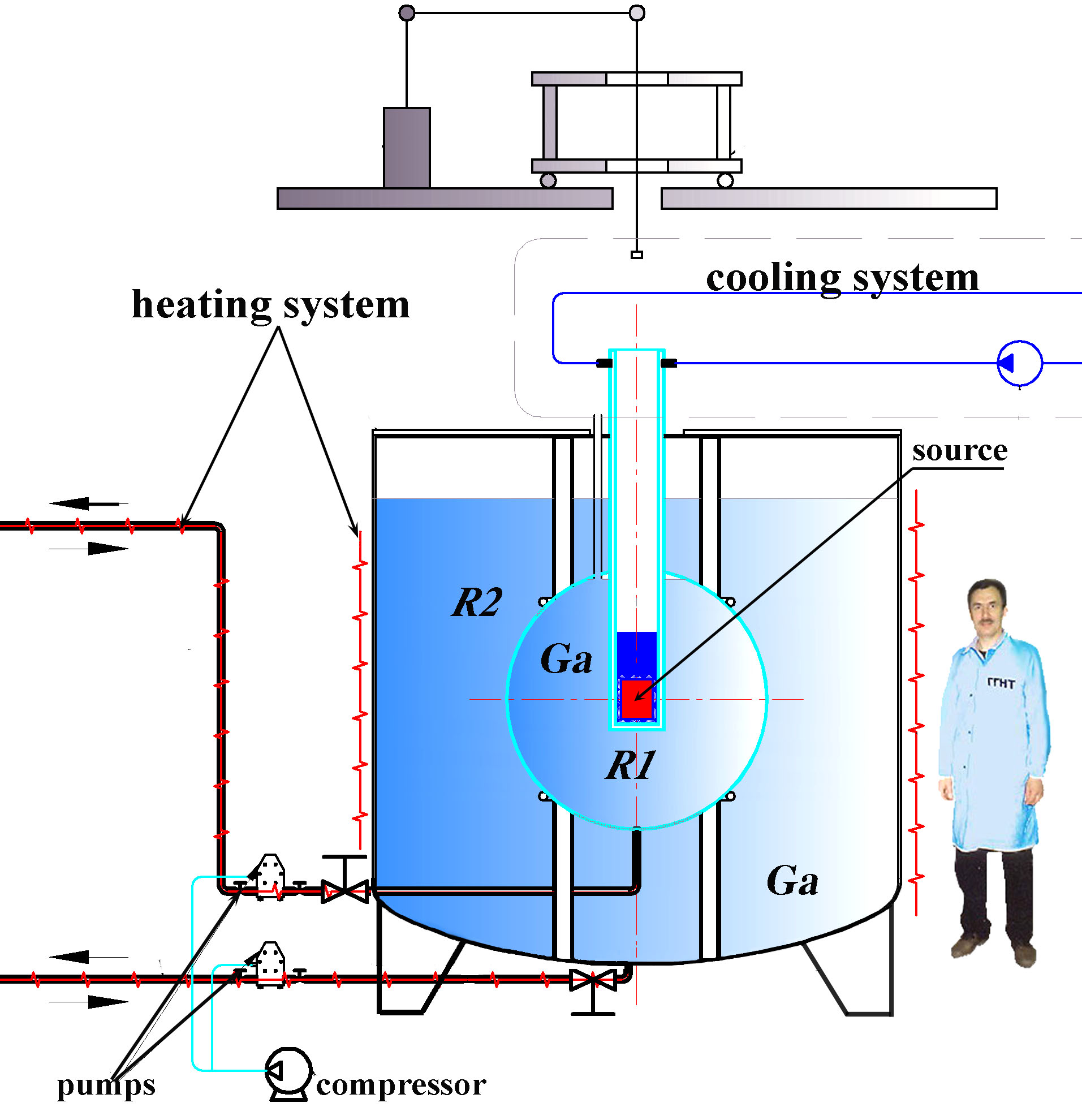}
\caption{BEST layout, vessel sizes are 
$R1=r^{\text{BEST}}_1$, $R2=r^{\text{BEST}}_2$.  
\label{BEST-layout}
}
\end{figure}

The source is a solid homogeneous cylinder with diameter about 9\,cm 
and height about 10\,cm. It is
placed in the center of sphere of radius $r^{\text{BEST}}_1=0.66$\,m 
filled with homogeneous liquid Gallium
${}^{71}$Ga. The sphere is inside the cylindrical vessel 
of radius $r^{\text{BEST}}_2=1.096$\,m and height $2\times r^{\text{BEST}}_2$ 
also filled with homogeneous liquid
Gallium, see Fig.\,\ref{BEST-layout}. 
The allocated for the artificial source central part of construction 
can be approximated\,\footnote{The use of the effective geometry changes the
  neutrino count rate by less than 2\%\,\cite{Abazajian:2012ys}, 
which impact on our estimates
  of the BEST sensitivity is negligible.} as 
a spherical region of radius
$r_c^{\text{BEST}}=10.5$\,cm. Only this part is free of Gallium. 
The electron neutrinos can be captured by ${}^{71}$Ga nuclei, which 
turn into ${}^{71}$Ge. This Germanium isotop decays solely by electron
capture to the ground state of  ${}^{71}$Ga with 
half-life of 11.43\,d. Without oscillations to the sterile neutrinos,
at the beginning of irradiation the mean production rate of ${}^{71}$Ge 
in each zone is 65 atoms per day. After an exposure period, 
the ${}^{71}$Ge atoms produced by neutrino capture are extracted
from Gallium and counted separately for each vessel with mostly the
same technique \cite{Abdurashitov:2011zz} as used for the SAGE
experiment. The lifetime of ${}^{51}$Cr is 27.7\,d and several
subsequent extractions are planned. A Monte Carlo simulation of the
entire experiment---10 extractions each with 9-day exposure---which
uses typical values of extraction efficiency, counter efficiency,
counter background rates, and includes the known solar neutrino rate,
indicates that the rate in each zone can be measured with a
statistical uncertainty of about $3.7\%$. An expected total systematic 
uncertainty is about $2.7\%$. The combined fit to the 
10 extractions enables measurement of the electron 
neutrino flux keeping it under control 
in each vessel with accuracy better 
than 5\%\,\cite{Abazajian:2012ys,Gavrin:2015aca}.

\section{Gallium anomaly}
\label{sec:Anomaly}

We start with analysis of anomalous results, observed in SAGE and
GALLEX, to find the region in sterile neutrino parameter space
$(\Delta m^2,\,\sin^22\theta)$ favored by the explanation of the
anomaly as oscillations into the sterile neutrinos. 

In SAGE the artificial sources have been placed in the center of 
spherical vessel, which for our purposes may be approximated as a
sphere of radius $r^{\text{SAGE}}=72.6$\,cm with central part of
radius $r_c^{\text{SAGE}}=25.3$\,cm free of Gallium and allocated for
the source. The artificial sources 
were of the cylindrical form with height of 15\,cm and diameter
of 9.5\,cm 
\cite{Abdurashitov:1998ne}. 
It can be approximated as a sphere of  
radius $r_s^{\text{SAGE}}=6.3$\,cm.  
 In the first experiment the ${}^{51}$Cr source 
was used. It provided the same neutrino spectrum as in case of BEST,
which is effectively two-peak with energy lines at  
$E_1=0.75$\,MeV (dominant peak) and $E_2=0.43$\,MeV. In the second
experiment the ${}^{37}$Ar source was used. The dominant neutrino 
mode is at $E_1=0.811$\,MeV, the subdominant is very close, 
$E_2=0.813$\,MeV, so the source is monochromatic with a high accuracy.

Neutrino flux at the distance $r$ from the source is proportional to
the survival probability \eqref{survival}. The rate of induced
transitions ${}^{71}\text{Ga}\to{}^{71}$Ge is also proportional  
to the Gallium density, which was uniform, and to the 
neutrino capture cross section, $\sigma_{\text{Ga}}(E)$, which is
different for different neutrino energies, and has been recently refined
\cite{Frekers:2011zz,Frekers:2013hsa,Frekers:2015wga}. 
The contribution of each neutrino line to the number of transition is
weighted with intensity of the line and neutrino caption cross section
$\sigma_{\text{Ga}}(E_i)$, $i=1,2$. 
Finally, for the ratio of signal,  
expected within the sterile neutrino hypothesis, and signal, expected
without sterile neutrino, we obtain
\begin{equation}
\label{theory}
R^{th}\!=\!\frac{1}{\Delta L}\!\!\int_{r_1}^{r_2} \!\!\!\!\!\!\! dr\! 
\left[ P(E_1,|\vec{r}\!-\!\delta \vec{r}|)f_1 + 
P(E_2,|\vec{r}\!-\!\delta \vec{r}|)f_2\right],
\end{equation}
where the relative contributions of the two lines in the ${}^{51}$Cr
source are $f_1=0.96$, $f_2=0.04$, the integration goes between the 
radius of the central part $r_1=r_c^{\text{SAGE}}$ and the vessel radius
$r_2=r^{\text{SAGE}}$, the normalizing effective length is
$\Delta L=r_2-r_1$; the results are further averaged over the
artificial source of 
finite size (variable $\delta {\vec r}$) adopting the spherical approximation
with radius  $r^{\text{SAGE}}_s$.

Likewise this formula can be applied to describe the results of the SAGE
experiment with ${}^{37}$Ar source. 

The two measurements in GALLEX experiment\,\cite{Cribier:1996cq} 
have been performed with ${}^{51}$Cr radioactive sources. 
In each case the radioactive source can be approximated by a homogeneous
sphere of radius $r_s^{\text{GALLEX}}\approx 0.4$\,m. 
The source was placed in a 
center of the vessel which for our purposes can be approximate as a
sphere of radius $r^{\text{GALLEX}}\approx 2.5$\,m, the radius of the
central part with source is $r_c^{\text{GALLEX}}\approx 0.45$\,m.  
To describe the expected ratio of 
the signals with sterile neutrino and without sterile neutrino one can
use the same formula \eqref{theory} with the same values of $E_i$ and 
$f_i$, as those adopted for the SAGE with ${}^{51}$Cr source, and 
$r_1=r_c^{\text{GALLEX}}$, $r_2=r^{\text{GALLEX}}$. 

The anomalous results read\,\cite{Giunti:2010zu}
\begin{equation}
\label{anomaly}
\begin{split}
R^{obs}_{\text{SAGE}}\l{}^{51}\text{Cr}\r&=0.95\pm0.12 \,,\;\;\;\\
R^{obs}_{\text{SAGE}}\l{}^{37}\text{Ar}\r&=0.79\pm0.10 \,,\;\;\;\\
R^{obs}_{\text{GALLEX}}\l{}^{51}\text{Cr}\r&=0.953\pm0.11 \,,\\
R^{obs}_{\text{GALLEX}}\l{}^{51}\text{Cr}\r&=0.812\pm0.11 \,.
\end{split}
\end{equation}
The best fit values of the sterile neutrino model parameters $(\Delta
m^2,\,\sin^22\theta)$ entering the theoretical expectations
\eqref{theory} through the survival probability \eqref{survival}, can
be obtained by minimizing the $\chi^2$ statistics defined as the
following sum over all the four experiments
\begin{equation}
\label{chi2}
\chi^2=\sum_{i=1}^4\frac{\l R_i^{obs}-R_i^{th}(\Delta
  m^2,\,\sin^22\theta) \r^2}{\sigma_i^2}\,,
\end{equation}  
where $\sigma_i$ stand for corresponding uncertainties 
in measured values \eqref{anomaly}. 

\begin{figure}[!htb]
\centering
\includegraphics[width=0.8\columnwidth]{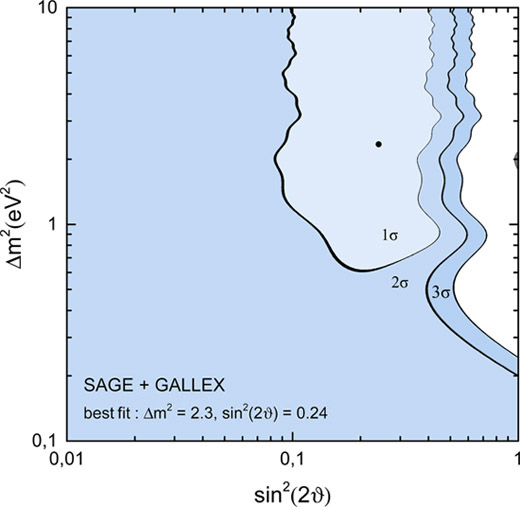}
\caption{Regions, obtained from the combined analysis of the results
  of the four radioactive source experiments by SAGE and
  GALLEX. The best fit value is
  \eqref{SG-best-fit}. 
\label{SAGE+GALLEX}
}
\end{figure}

We find for the best-fit value 
\begin{equation}
\label{SG-best-fit}
\Delta m^2\approx 2.3\,\text{eV}^2\,,\;\;\;\;\sin^2\theta\approx0.24\,,
\end{equation} 
which we exploit below as a refined version of the estimate
\eqref{best-fit-region}.  In Fig.\,\ref{SAGE+GALLEX} 
we present the contours outlining the regions consistent with the sterile
neutrino explanation of the anomaly at 1\,$\sigma$, 2\,$\sigma$ and
3\,$\sigma$ confidence levels. The contours refer to the 
corresponding marginal values of $\chi^2=\chi^2_{min}+\Delta\chi^2$ 
for $\Delta\chi^2$ with two free parameters:
$\Delta\chi^2=2.30,\,6.18,\,11.83$, respectively \cite{Agashe:2014kda};
the oscillation parameters $\Delta m^2$, $\sin^22\theta$ are 
adopted in the minimized $\chi^2$-function for the four Gallium 
radioactive source experiments. As one observes
from Fig.\,\ref{SAGE+GALLEX}, the lines of constant 
$\chi^2$ forms
rather shallow profile, so the best-fit value \eqref{SG-best-fit} is 
not actually indicative: the 1\,$\sigma$-region is quite broad,
$\Delta m^2\gtrsim 0.7$\,eV$^2$, $0.1\lesssim \sin^22\theta\lesssim
0.4$. It is in agreement, of course, with the 
statement\,\cite{Giunti:2010zu}  that the
statistical significance of the observed anomaly is not high, at the
level of 2-3\,$\sigma$, as we mention in Introduction.

\section{Sensitivity contours}
\label{sec:Results}

As noticed in sec.\,\ref{sec:Exp} the technique used in BEST
allows to measure the ratio $R$ in each vessel with accuracy of about
$\sigma_{\text{BEST}}=0.05\times R$. The presence of two vessels enables 
performing two independent 
experiments in the same time. The predictions for the expected ratios
are given by the same formula \eqref{theory} with
$r_1=r_c^{\text{BEST}}$ and $r_2=r_1^{\text{BEST}}$ for experiment one and
$r_1=r_1^{\text{BEST}}$ and $r_2=r_2^{\text{BEST}}$ for experiment two. 
  
Further we treat the would-be measured at BEST ratio $R_m$ as a
random Gaussian variable with central value $R_c$ and standard deviation
$\sigma_{\text{BEST}}=0.05\times R_c$, so that $R_m$ is distributed as 
\begin{equation}
\label{Ddist}
D_{R_c,\sigma_{\text{BEST}}}(R_m)=\frac{1}{\sqrt{2\pi}\sigma_{\text{BEST}}} 
\exp\l-\frac{\l R_m-R_c\r^2}{2\sigma_{\text{BEST}}^2}\r.
\end{equation}
For each value of $R_m$ the corresponding model parameters are
determined by maximization of the likelihood function 
\begin{equation}
\label{chi2-BEST}
\begin{split}
{\cal L}_{\sigma_{\text{BEST}}}(\Delta m^2,\sin^22\theta;R_m)\\\propto \exp
\l -\frac{\l R_m-R^{th}(\Delta m^2,\sin^22\theta)\r^2}{2\sigma^2_{\text{BEST}}}\r
\end{split}
\end{equation}
\begin{figure}[!htb]
\centering
\includegraphics[width=0.85\columnwidth]{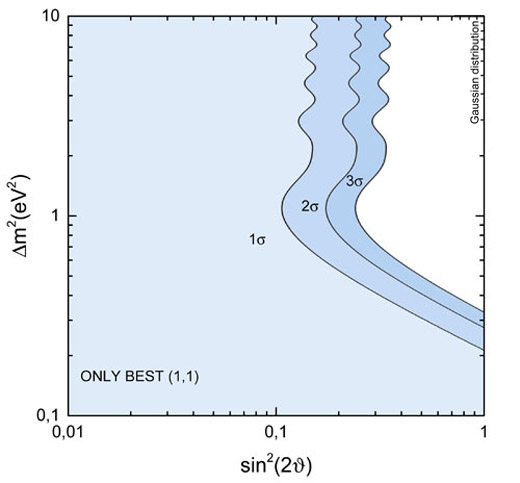}

\includegraphics[width=0.85\columnwidth]{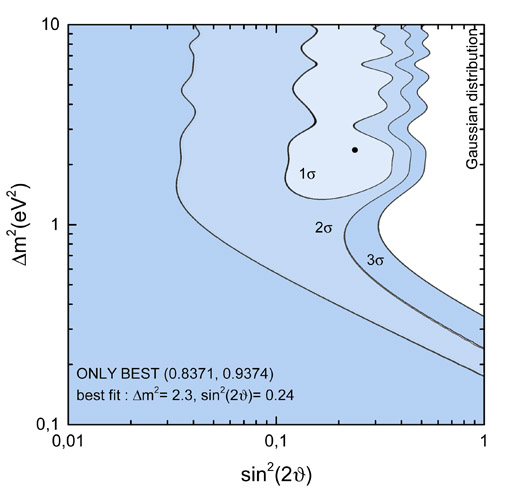}
\caption{Regions, favored by the BEST experiment 
in cases: (upper panel) it finds no anomaly, (lower
  panel) it confirms the Gallium anomaly; see the main text for details. 
\label{BEST-sensitivity}
}
\end{figure}
Suppose, that the BEST results are consistent with a particular 
theory prediction $R_c$. Then within the Bayesian approach 
the favored value of the sterile
neutrino parameters are determined from \eqref{chi2-BEST}  
marginalized over $R_m$ with Gaussian prior \eqref{Ddist}, 
which gives 
\begin{align*} 
\int dR_m &{\cal L}_{\sigma_{\text{BEST}}}(\Delta
   m^2,\sin^22\theta;R_m)D_{R_c,\sigma_{\text{BEST}}}(R_m)\\
&{=\cal L}_{\sqrt{2}\sigma_{\text{BEST}}}(\Delta
   m^2,\sin^22\theta;R_c).
\end{align*}
Thus, the favored model parameters are
distributed as \eqref{chi2-BEST} with the following replacement $R_m\to
R_c$ and $\sigma_{\text{BEST}}\to\sqrt{2}\sigma_{\text{BEST}}$. 

Then the BEST sensitivity can be obtained using the same
$\chi^2$-expression \eqref{chi2} with $R_i^{obs}=R_c$ and
$\sigma_i=\sqrt{2} \sigma_{\text{BEST}}$. To illustrate this formula 
in Fig.\,\ref{BEST-sensitivity} we present the plots with exclusion
regions for two cases: (top panel) BEST observations are fully
consistent with only three neutrino states, i.e. $R_c=1$ for each
vessel; (bottom panel) BEST observations are fully consistent with
the best fit oscillation parameters \eqref{SG-best-fit} from the 
sterile neutrino explanation of the Gallium anomaly, 
which for the BEST setup implies  
\begin{equation}
\label{confirmation}
R_{c,1}=0.8371\,,\;\;\;R_{c,2}=0.9374
\end{equation}
 for the inner and outer vessels, respectively. 
The regions outside the
2\,$\sigma$        contours will be excluded by BEST in these two
cases at 95\%CL, which illustrate the BEST prospects in
testing the Gallium anomaly. 
\begin{figure}[!htb]
\centering 
\includegraphics[width=0.85\columnwidth]{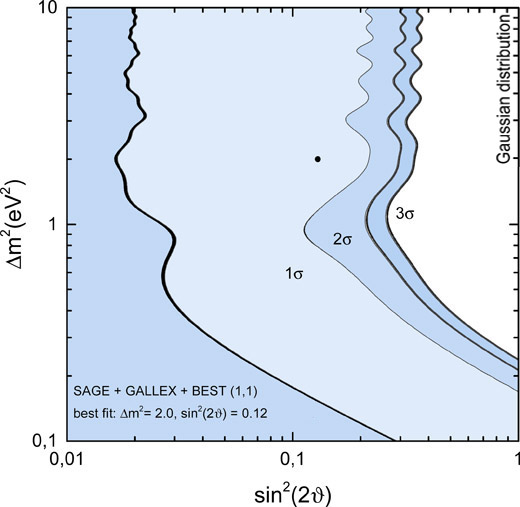}

\includegraphics[width=0.85\columnwidth]{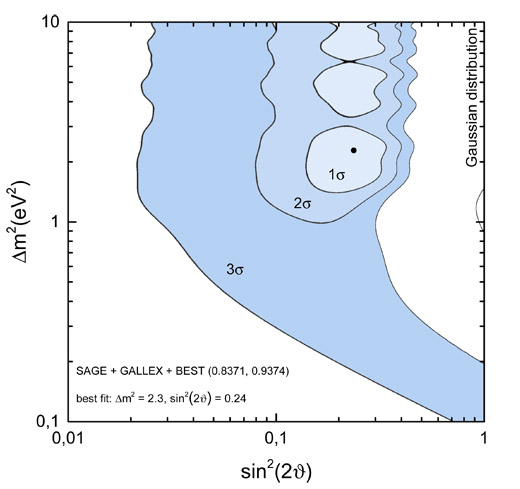}
\caption{Regions, favored by the combined analysis of all the Gallium 
radioactive source  experiments in cases: 
(upper panel) BEST finds no anomaly, (lower
  panel) BEST confirms the Gallium anomaly; see the main text for details. 
\label{all-6}
}
\end{figure}

The plots in Fig.\,\ref{BEST-sensitivity}, as compared to that in
Fig.\,\ref{SAGE+GALLEX},  ensure that the status of
the Gallium anomaly after the BEST experiment will largely depend on
its results. To support this conclusion, we calculate the joint
$\chi^2$-statistics \eqref{chi2} for all the radioactive source 
experiments by SAGE,
GALLEX, and BEST (6 experiments in total) assuming for the latter
an error $\sigma_i=\sqrt{2} \sigma_{\text{BEST}}$ and central values 
$R_{i}^{obs}$ consistent with either the three neutrinos only, hence 
$R_{c,1}=R_{c,2}=1$,  
or with the best-fit values \eqref{SG-best-fit}, and hence ratios   
\eqref{confirmation}. The favored in these cases regions of the model
parameter space are presented in Fig.\,\ref{all-6}.

\section{Conclusions}
\label{sec:Concl}

To summarize, we present the refined estimates of BEST sensitivity to
models with light sterile neutrinos mixed with electron neutrinos. We
study possible impact of the future BEST results on the status of the
Gallium anomaly. 

The performed numerical studies  are illustrated with plots in
Figs.\,\ref{BEST-sensitivity}, \ref{all-6}. It is worth noting that 
the region $\Delta m^2\gtrsim 100$\,eV$^2$ is excluded for
$\sin^22\theta>0.1$ by the peak searches in $\beta$-decays, the
strongest limits are placed by the Troitsk $\nu$-mass
experiment\,\cite{Belesev:2012hx,Belesev:2013cba}. Note also, 
that sterile neutrino of mass $m_s\simeq 1$\,eV noticeably changes the
cosmological prediction of the standard $\Lambda$CDM model. 
In particular, the Planck experiment \cite{Ade:2015xua} 
excludes masses above $0.5$\,eV for the fully thermalized sterile
neutrino. However, this cosmological 
limit considerably depends on the cosmological data set
used in the analysis. Also, this limit is inapplicable if the sterile
neutrinos are not thermalized in the early Universe 
plasma of the Standard Model particles, which can happen in specific
extensions of the Standard Model, see 
e.g.\,\cite{Hannestad:2013ana,Dasgupta:2013zpn}. Thus, cosmology still
allows for the presence of light sterile neutrinos (introduced to explain
the Gallium anomaly) with extended particle physics and/or
cosmological model. In turn, direct searches for light sterile
neutrinos, like that provided by BEST, can test such extensions.

We thanks F.\,Bezrukov, S.\,Demidov and G.\,Rubtsov for valuable
discussions. 
The work was supported by the RSF grant 14-22-00161.

\bibliographystyle{apsrev4-1}
\bibliography{Sensitivity}

\end{document}